\documentclass[letterpaper]{article} 
\usepackage{aaai25}  
\usepackage{times}  
\usepackage{helvet}  
\usepackage{courier}  
\usepackage[hyphens]{url}  
\usepackage{graphicx} 
\usepackage{natbib}  
\usepackage{caption} 
\usepackage{booktabs} 
\usepackage{amsmath}
\usepackage{amssymb}
\usepackage{amsthm}
\newtheorem{definition}{Definition}

\usepackage{algorithm}
\usepackage{algorithmic}

\usepackage{newfloat}
\usepackage{listings}
\DeclareCaptionStyle{ruled}{labelfont=normalfont,labelsep=colon,strut=off} 
\floatstyle{ruled}
\newfloat{listing}{tb}{lst}{}
\floatname{listing}{Listing}
\title{GTDE: Grouped Training with Decentralized Execution for Multi-agent Actor-Critic}
\author{
    Mengxian Li\textsuperscript{\rm 1,\rm 2}, Qi Wang\textsuperscript{\rm 1}\thanks{corresponding author}, Yongjun Xu\textsuperscript{\rm 1} 
}
\affiliations{
    \textsuperscript{\rm 1}Institute of Compute Technology, Chinese Academy of Sciences\\
    \textsuperscript{\rm 2}University of Chinese Academy of Sciences\\

    limengxian24b@ict.ac.cn, wangqi08@ict.ac.cn
}

\usepackage{bibentry}

\begin{document}

\maketitle

\begin{abstract}
The rapid advancement of multi-agent reinforcement learning (MARL) has given rise to diverse training paradigms to learn the policies of each agent in the multi-agent system. The paradigms of decentralized training and execution (DTDE) and centralized training with decentralized execution (CTDE) have been proposed and widely applied. However, as the number of agents increases, the inherent limitations of these frameworks significantly degrade the performance metrics, such as win rate, total reward, etc. To reduce the influence of the increasing number of agents on the performance metrics, we propose a novel training paradigm of grouped training decentralized execution (GTDE). This framework eliminates the need for a centralized module and relies solely on local information, effectively meeting the training requirements of large-scale multi-agent systems. Specifically, we first introduce an adaptive grouping module, which divides each agent into different groups based on their observation history. To implement end-to-end training, GTDE uses Gumbel-Sigmoid for efficient point-to-point sampling on the grouping distribution while ensuring gradient backpropagation. To adapt to the uncertainty in the number of members in a group, two methods are used to implement a group information aggregation module that merges member information within the group. Empirical results show that in a cooperative environment with 495 agents, GTDE increased the total reward by an average of 382\% compared to the baseline. In a competitive environment with 64 agents, GTDE achieved a 100\% win rate against the baseline.
\end{abstract}

\begin{links}
    \link{Code}{https://github.com/lemonsinx/GTDE}
\end{links}

%

\section{Introduction} \label{sec:intro}
Multi-agent reinforcement learning has recently made remarkable progress across diverse real-world applications, such as multi-robot cooperative tasks \cite{MRW}, complex games \cite{game1, game2}, computer network \cite{CN1, CN2}, and autonomous driving \cite{ad1, ad2}. These scenarios are often constrained by partial observability or limited communication, necessitating the use of decentralized policies. These policies rely solely on the observation history of each individual agent to make decisions. The decentralized policies also avoid the issue of exponential growth in the action space as the number of agents increases, making a fully centralized policy impractical for large-scale multi-agent systems. The decentralized policies can be learned through either Decentralized Training and Execution (DTDE) and Centralized Training with Decentralized Execution (CTDE) \cite{marl-book}. 

DTDE utilizes a decentralized approach to simplify the training of multi-agents by using a single-agent RL algorithm for each individual agent. However, this training paradigm introduces two inevitable issues. Firstly, agents are unable to leverage information from other agents in the environment, making cooperation difficult \cite{MANSA}. Secondly, since each agent treats other agents as part of the environment, they are affected by environmental non-stationary caused by updates in the policies of other agents \cite{IL_not_markovian_2011}, which intensifies with a larger number of agents.

CTDE uses a centralized approach to train decentralized policies \cite{CTDE_2008, CTDE_2016}, allowing agents to utilize global information during training while effectively avoiding the non-stationarity of the environment. However, due to limitations in partial observability, obtaining comprehensive global information is often infeasible. Fortunately, we can replace global information with joint information. Despite this, as the number of agents increases, three challenges remain. Firstly, the use of joint information faces the curse of dimensionality \cite{MADRL2023}, which makes it impossible to train in large-scale scenarios, such as the Gather scenario with 495 agents \cite{MAgent2}. Secondly, in regions where agents have weak interaction, joint information may not offer significant benefits \cite{scq}, indicating potential information redundancy. Thirdly, communication restrictions may prevent each agent from obtaining joint observations from all other agents, which poses a challenge for CTDE paradigms to train in real-world applications.

\begin{figure}[t]
    \centering
    \includegraphics[width=0.9\columnwidth]{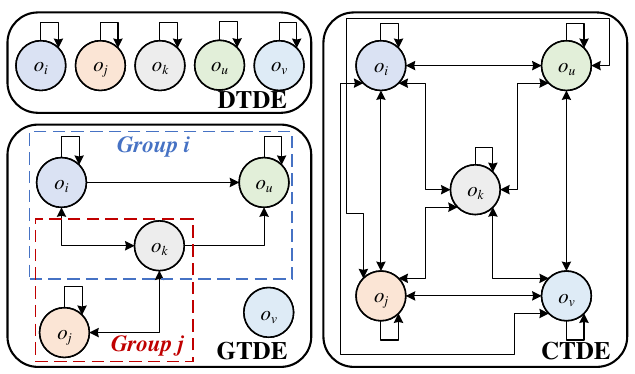} 
    \caption{Each circle represents the observation of an agent, and a unidirectional arrow such as $o_i\rightarrow o_k$ indicates that agent $i$ needs the observation of agent $k$ during training.}
    \label{fig:graphlink}
\end{figure}

With an increasing number of agents, the aforementioned two paradigms may no longer apply to the training of policies. To address the training problem of large-scale multi-agent systems, we consider abstracting the information required by agents during the training of policies as directed links between agents, thereby obtaining a directed graph. Through this approach, we can obtain that DTDE using individual information\footnote{We utilize "individual information" to denote the information of an agent, "joint information" to represent the information of all agents, and "local information" to refer to the information of partial agents.} represents directed links between agents as self-linked, while CTDE using joint information represents directed links between agents as pairwise links. There is a situation between these two frameworks, as shown in Fig.\ref{fig:graphlink}, which is the partial links between agents. This also indicates a fact that agents should focus on the specific local information they require \cite{MAAC}, rather than relying on individual or joint information from agents. Using local information for training of policies can effectively alleviate the problems in DTDE and CTDE, and also provide the possibility for large-scale multi-agent training in real-world applications. For the sake of simplicity, we define the information required by an agent as a group of this agent, as described in Def.\ref{de:group}. 

In this paper, we propose a novel framework called Grouped Training with Decentralized Execution (GTDE). GTDE mainly considers dividing a large-scale multi-agent system into numerous small groups during training at each timestep, and each agent only needs to utilize the information within the group. However, the dynamic nature of the environment presents an issue in that the information required by each agent can vary over time. Therefore, we propose an adaptive grouping module, in which the number of groups and the number of members within the same group can vary continuously over time. At each timestep, adaptive grouping is only based on observation history to obtain the multivariate Bernoulli distribution of each agent link, and Gumbel-Sigmoid (a variant of Gumbel-Softmax) is used to sample the distribution. After that, given the uncertainty regarding the number of members within the group, we use two methods to implement the group information aggregation module to efficiently consolidate information within the group. During the execution of policies, agents take actions only based on their individual information, such as DTDE and CTDE. Because value-based methods require the same information to be provided for a single Q-value during training and execution, GTDE primarily focuses on extending multi-agent policy gradient methods within its framework, i.e. actor-critic algorithm. Our contributions are as follows:
\begin{itemize}
   \item We propose a framework of GTDE oriented towards information requirements for partially observable large-scale multi-agent systems, which focuses on the training of agents from local information rather than global and joint (CTDE) information or individual (DTDE) information. 
   \item We present an adaptive grouping method that combines Gumbel-Sigmoid to address the dynamic information requirements of agents and use group information aggregation to consolidate information within the group. 
   \item We evaluate GTDE on the StarCraft Multi-Agent Challenge V2 (SMACv2) \cite{SMACV2} with 20 agents, the Battle scenario with 64 agents, and the Gather scenario with 495 agents \cite{MAgent}. Experiments have demonstrated that with an increasing number of agents, GTDE outperforms both DTDE and CTDE. 
\end{itemize}

\section{Related Works} \label{sec:RW}
Several algorithms \cite{MADDPG, MAA2C, IL_2020, MAPPO} based on CTDE and DTDE encounter challenges such as environmental non-stationarity and the curse of dimensionality, as mentioned in Sec.\ref{sec:intro}. MAAC \cite{MAAC} and G2Anet \cite{G2Anet} identified these issues and addressed them by introducing a soft/hard attention mechanism, which can effectively address the challenges of dimensionality and information redundancy in the centralized critic. However, this approach introduces a new issue. Centralized attention mechanisms necessitate the utilization of information from all agents, making it impossible to train in large-scale scenarios. Therefore, we consider solving the above problem by grouping without introducing other centralized modules. 

Some early works utilized the prior domain knowledge to group the agents \cite{agent_grouping_2002, agent_grouping_2013}. Mean-Field \cite{MF} RL uses the actions of nearby agents to estimate the Q-function, but this method does not explicitly define how to determine the neighbors of agents. Subsequently, role-based \cite{ROMA, ROGC} and subtask-based \cite{LDSA, MACC, QSCAN} methods group agents to different roles or tasks, and agents of the same role or subtask are naturally grouped together. However, subtask-based and role-based methods require the introduction of an additional centralized module for grouping, and some of them need to perform a clustering algorithm at each time step before running the MARL algorithm, incurring a high time cost. Other works, such as REFIL \cite{REFIL} divides agents into two groups, VAST \cite{VAST} uses meta-policy for agent grouping, DHCG \cite{DHCG} learning grouping policy using RL method, and CommFormer \cite{CommFormer} learning a static graph. The above grouping algorithms all have the same problem, which is that the number of groups or the number of members within the same group is fixed, which means that the dynamic nature of the environment has not been fully considered. As the number of agents increases, they need to specify specific grouping numbers for different environments, which is not flexible. 

In our method, observation history from a single agent is used to determine the group, and training is conducted end-to-end. Allow agents to learn the number of groups and the number of members within a group by themselves rather than artificially specifying. The model of sharing information during the execution of policies is complementary to GTDE and can combine the two approaches to generate further performance advantages.

\section{Background} \label{sec:bg}
\subsection{Decentralized Partially Observable Markov Decision Processes (DEC-POMDP)}
We study DEC-POMDP \cite{DEC_POMDP}, which is defined by a tuple $\left \langle  n, S, Z, O, A, R, P, \gamma \right \rangle$, where $n$ is the number of agents. $s \in S$ is the true state of the environment. At every stage $t$, the environment obtains joint observations $z\in Z$ through the observation function $O(s^t, a^{t - 1})$ and emits them to each agent. Each agent $i$ has an observation history $\tau_i^t\equiv(o^1_i,\dots, o^t_i)$\footnote{We use $o_i$ to denote the observation of the i-th agent and $o^t$ to denote the observation at the t-th timestep.}. Each agent $i$ selects action $a_i\in A_i$ based on a parameterized policy $\pi_i(a_i|\tau_i;\theta)$. The environment transitions to the next state based on the joint actions $a\in A$ of all agents and state transition functions $P(s^{t + 1}|s^t, a^t)$, giving the same reward $r^{t+1}\in R(s^t, a^t)$ to all agents through the reward function $R(s^t, a^t)$. The objective of the agents is to maximize the expected return $J=\sum_{t=1}^T\gamma^{t-1}r^t$ where $\gamma\in[0, 1)$ is a discount factor and $T$ is the episode length. 

\subsection{Actor-Critic Algorithms}
The Policy Gradient (PG) method \cite{RL2nd, PG} directly learns parameterized policies to maximize the expected return of the agent, with the optimization objective $J(\theta)=\mathbb E_{s\sim p_\pi, a\sim \pi(\theta)}[R(s, a)]$. The gradient with respect to the policy parameters is denoted as 
\begin{equation*}
   \nabla_\theta J(\theta)=
   \mathbb E_{s\sim p_\pi, a\sim \pi(\theta)}[\nabla_\theta \log{\pi(a|\tau;\theta)}Q_\pi(\tau,a; \varphi)]
\end{equation*}
where $p_\pi$ is the state distribution under policy $\pi$. Actor-Critic Algorithms \cite{AC_2000} approximates state-action value $Q_\pi(s,a)$ through TD learning. It also uses parameterized functions for updates 
\begin{equation}
   \mathcal L_{TD}(\varphi)= \mathbb E_\mathcal D[(Q^t - (R(s^t, a^t) + \gamma Q^{t+1}))^2]
   \label{eq:Qv}
\end{equation}
where $\mathcal D$ is a set of tuples $(s^t, a^t, r^{t+1}, s^{t+1})$ generated through interaction with the environment and $Q^t=Q(\tau^t, a^t; \varphi)$.

\subsection{Multi-Agent Actor-Critic Algorithms}
In the framework of CTDE, the multi-agent actor-critic algorithm utilizes the joint information of each agent to learn a critic for each agent or directly learn a centralized critic. The gradient of the expected return $J(\theta_i)$ with respect to the policy parameters is approximated by 
\begin{equation*}
   \nabla_{\theta_{i}} J(\theta_i)=
   \mathbb E_\mathcal D{[\nabla_{\theta_{i}} \log{\pi(a_i|\tau_i;\theta_i)}Q_\pi(\tau,a;\varphi_i)]}
\end{equation*} 
In the framework of DTDE, the multi-agent actor-critic algorithm only utilizes individual information of each agent to learn a critic for each agent 
\begin{equation*}
   \nabla_{\theta_{i}} J(\theta_i)=
   \mathbb E_\mathcal D{[\nabla_{\theta_{i}} \log{\pi(a_i|\tau_i;\theta_i)}Q_\pi(\tau_i,a_i;\varphi_i)]}
\end{equation*}
This paper also uses the PPO algorithm and therefore provides the PPO algorithm of DTDE and CTDE paradigms. The policy is updated to maximize  
\begin{equation}
    J(\theta_i)=
    \mathbb E_\mathcal D{[\min(\frac{\pi_{new}}{\pi}
    A_\pi, clip(\frac{\pi_{new}}{\pi}, 
    1\pm\epsilon)A_\pi)]}
    \label{eq:ppo}
\end{equation}
where $clip(\dot, 1\pm\epsilon)$ operator clips the input to $1-\epsilon / 1+\epsilon$ if it is below/above this value, and $\pi=\pi(a_i|\tau_i;\theta_i)$. For the IPPO algorithm, $A_\pi=A_\pi(\tau_i,a_i;\varphi_i)$ and for the MAPPO algorithm, $A_\pi=A_\pi(\tau,a;\varphi_i)$.

\section{Methods} 
We abstract the information required between agents into a directed graph $G^t=(V^t, E^t)$, where $G^t$ represents the directed graph at timestep $t$, $V^t$ is the set of nodes representing the agents and $E^t$ is the set of edges between nodes at timestep $t$ representing the information required by the agent. It can be seen that due to the dynamism of multi-agent systems, their directed graphs are inconsistent at each timestep. 
\begin{definition} \label{de:group}
    For the $i$-th node $v_i$ in a directed graph, we set the group of $v_i$ to $g(v_i)$, then
    \begin{equation}
        g(v_i)=\{c|I_{E^t}(<v_i, c>)=1, c\in V^t\} \label{eq:group}
    \end{equation}
\end{definition}
where $I_{E^t}(<v_i, c>)=\begin{cases} 1, <v_i, c>\in E^t \\ 0, <v_i, c>\notin E^t\end{cases}$, and $<u, v>$ represents the directed edge from node $u$ to node $v$. 

\subsection{GTDE} 
\begin{figure*}[!t]
    \centering
    \includegraphics[width=0.8\textwidth]{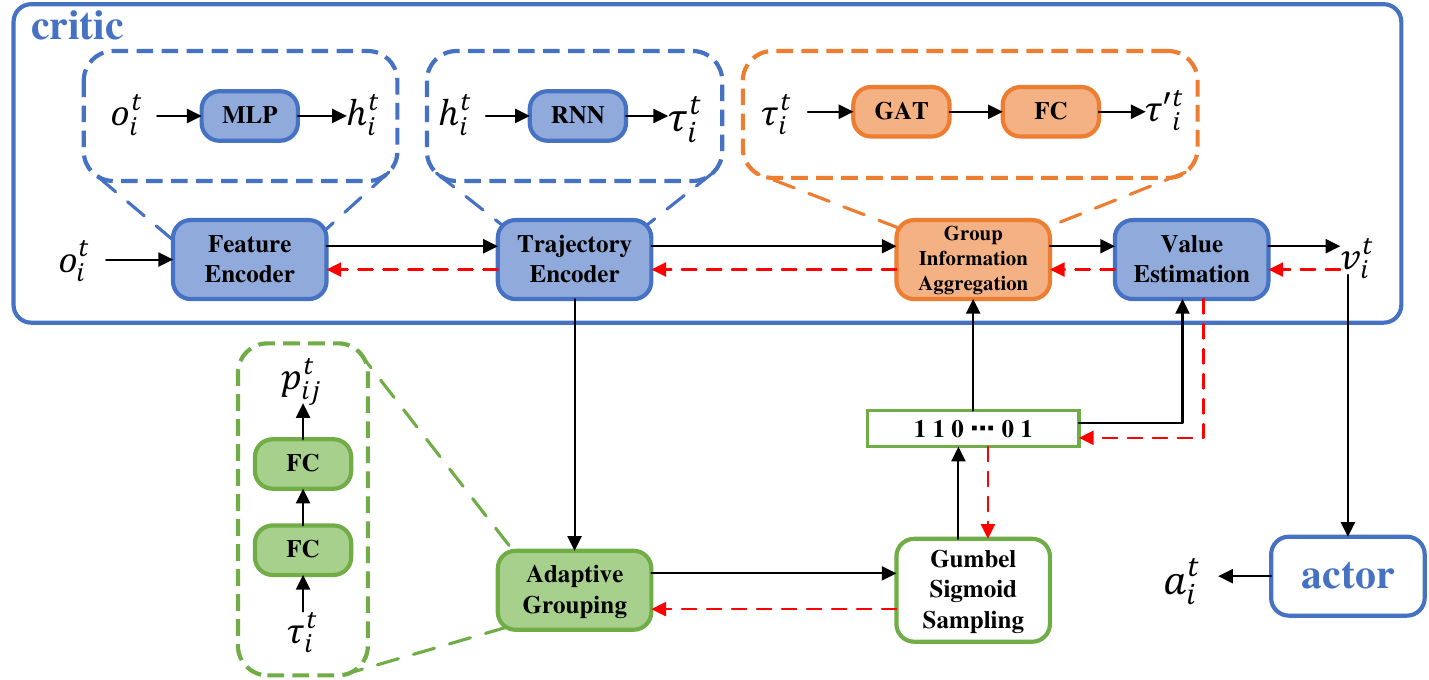}
    \caption{Overview of GTDE framework. The red dashed line represents gradient flow.}
    \label{fig:GTDEframe}
\end{figure*}

As shown in Fig.\ref{fig:GTDEframe}, the GTDE framework extends the DTDE and CTDE framework by introducing two additional components: the adaptive grouping module and the group information aggregation module. The adaptive grouping module enables the division of agents into distinct groups based on their individual observation history. The group information aggregation module fuses information from group members. Different from the previous grouping method, GTDE does not enforce a fixed number of groups or group members. Instead, it dynamically forms groups based on the observation history of each agent.

CTDE utilizes the joint information of the agent to estimate the value function, while DTDE relies on the individual information of the agent for the same purpose. In contrast, as shown in Fig.\ref{fig:graphlink}, GTDE employs partial information from the agent within the group to estimate the value function. We observed that when a multi-agent system with $n$ agents is represented as a directed graph, where nodes signify the observations of the agents and edges denote information required between agents. Notably, during the training of policies, the CTDE framework treats the entire multi-agent system as one group, where each group consists of $n$ members. In contrast, the DTDE framework views the multi-agent system as $n$ separate groups, assigning only one member to each group. GTDE framework takes an intermediate approach, categorizing agents into $g$ groups, with $m_g$ members in each group, where $1\leq g,m_g\leq n$. CTDE aligns with GTDE when all agents are pairwise linked, while DTDE is equivalent to GTDE when all agents are self-linked. One of the goals of GTDE is to predict the edges of links. Once the link is determined, the system is naturally divided into many groups. GTDE relies on intra-group information during training, rather than global or joint information, and only relies on individual information during execution, making it practical in real-world applications.

\subsection{Adaptive Grouping} \label{subsec:AG}
The adaptive grouping module determines the links of agent $i$ through $p_{ij}(c_i=j|\tau_i)$, where $\tau_i$ is the observation history of the agent $i$ and $c_i=j$ is the $i$-th agent is linked to the $j$-th agent. The agent only needs information from the other agents it is linked to, which avoids the challenge of assigning the agent to a specific group when the number of groups is uncertain. This probability distribution $p_{ij}(c_i=j|\tau_i)$ can be determined using physical distance, but introduces a new problem. Consider an environment with two buttons placed far apart, where pressing them simultaneously yields rewards. In this setting, with many agents present, only the two agents closest to the buttons are grouped together as the most appropriate choice. However, relying solely on physical distance assigns a very small probability of linking them. This also indicates that multi-agent grouping cannot be equated with data clustering. Even agents with highly unrelated features may belong to the same group. 

The adaptive grouping module predicts the edge set $E^t$ of the directed graph and uses this edge set to group multi-agent systems using Eq.\ref{eq:group}. Compared to prior-based fixed grouping methods, the adaptive grouping module dynamically adjusts the number of groups and members within the group at each timestep based on the dynamic changes in the environment. Meanwhile, due to the absence of centralized modules, each agent selects one or more links based on their own observation history during training. After all agents have completed grouping, a multi-agent system is modeled as a directed graph. Specifically, the adaptive grouping module is parameterized by a neural network, with the input being the observation history of a single agent and the output being the $n$-variate Bernoulli distribution, where $n$ is the number of all agents. Sample on the $n$-variable Bernoulli distribution to obtain the final determined grouping distribution. It should be noted that each row in the adjacency matrix $A^t$ of a directed graph $G^t$ represents the grouping distribution of an agent. According to the definition of the group, when the grouping distribution of agents is consistent, their belonging groups are also consistent, which makes the number of groups not fixed. At the same time, sampling from the $n$-variate Bernoulli distribution can obtain different grouping distribution, which makes the number of members in the same group not fixed. Due to the non-differentialable of the sampling operator in backpropagation, we need to use the reparameterization trick of discrete distributions. We can use Gumbel-Sigmoid to efficiently sample $n$-variable Bernoulli distribution, which is equivalent to Gumbel-Softmax. Please refer to Appendix \ref{ap:gs} for a detailed discussion. As depicted in Sec.\ref{sec:bg}, the $J(\theta)$ under the GTDE framework in the Actor-Critic algorithm is updated to maximize 
\begin{equation*}
    J(\theta)=
    \mathbb E_\mathcal D{[\log{\pi(a_i|\tau_i;\theta_i)}Q_\pi(\tau'_{g(v_i)},a_i;\varphi)]}
\end{equation*}
where $\tau'_{g(v_i)}$ is the aggregation information of observation history from agent $i$ and its links(To be mentioned in the next section). For the PPO algorithm, consistent with Eq.\ref{eq:ppo}, where $A_\pi=A_\pi(\tau'_{g(v_i)},a_i;\varphi)$. For simplicity, this paper employs parameter sharing, setting $\theta=\theta_1=\theta_2=\dots=\theta_n$.

\subsection{Group Information Aggregation} \label{subsec:GIA}
After the adaptive grouping module obtains the adjacency matrix of the directed graph $G^t$, the group information aggregation module can be used to merge the information of the linked agents. Each agent requires its own information, so we need to add a mask matrix and set the diagonal elements in the adjacency matrix to $1$ to ensure that each agent is self-linked. Then, we introduce two methods for aggregating group information. 

\subsubsection{Matrix Multiplication}
We multiply the adjacency matrix with the agent trajectory matrix, denoted as $\tau'_{g(v_i)}=A^t\tau^t$. Since the elements in the adjacency matrix are either $0$ or $1$, the result matrix $\tau'_{g(v_i)}$ only focuses on the trajectory information of linked agents, disregarding the trajectory information of unlinked agents. This implies that agent $i$ only requires the trajectory information of agent $j$ corresponding to the elements in the $i$-th row vector of the adjacency matrix that have a value of $1$. The $i$-th row in the result matrix is the sum of the information required by the $i$-th agent. By this means, agent $i$ no longer requires a centralized module to obtain information about all agents. Instead, it only needs to acquire information from a subset of agents to complete its learning process. The use of an adjacency matrix facilitates tensor computations under parameter sharing. In the case of non-parameter sharing, agent $i$ derives a row vector of grouped distributions from its own observation history and then retrieves the trajectory information of agent $j$ corresponding to the elements in the vector that have a value of 1, to obtain $Q_\pi$ or $A_\pi$.

\subsubsection{Graph Attention Network}
The $i$-th row of the result matrix can be written as $[\tau'_{g(v_i)}]_{i}=\sum_{j\;s.t.\;A_{ij}=1}\tau_{j}$. For some complex scenarios, we can assign different weights to trajectory information instead of giving all selected trajectories a weight of 1. This can be done by introducing a parameter, $\alpha$, before the trajectory, i.e. $[\tau'_{g(v_i)}]_{i}=\sum_{j\;s.t.\;A_{ij}=1}\alpha_{ij}\tau_{j}$. Graph attention network(GAT) \cite{GAT} can achieve this goal. Specifically, the attention weight $e_{ij}$ of node $i$ to node $j$ is $e_{ij}=w_1\tau_i+w_2\tau_j$. By using the above formula, we can obtain the attention matrix $e$. It is worth noting that although all agent information is used in calculating the attention matrix, unlinked nodes are masked based on the adjacency matrix. As a result, the attention weights of unlinked agents are irrelevant, meaning that any value in those positions is acceptable. Afterwards, we use the Softmax function to obtain the final attention weights, denoted as $\alpha_{ij}$. Furthermore, although GTDE can complete training with only local information, there are still some challenges in real-world applications. One of them is that although agent $i$ aims to obtain the trajectory information of agent $j$, the information may not be reached due to communication issues. To address this, we use a mask matrix to randomly drop values in the adjacency matrix with a probability of 0.1. This approach not only simulates the aforementioned communication issue but also prevents the agent from simultaneously selecting information from all agents, thereby avoiding a regression to the CTDE framework.

\section{Experiments} \label{sec:exp}
In this section, we compared GTDE with CTDE and DTDE and also evaluated variants of GTDE used for ablation study, represented as GTDE-F with fixed grouping distribution and GTDE-U with uniform grouping distribution, as well as GTDE-A with pairwise links, i.e. GTDE degenerated into CTDE. All algorithms were trained using an NVIDIA GeForce RTX 4090 GPU. 

\subsection{Experimental setup}
\subsubsection{SMACv2}
SMACv2, like SMAC \cite{SMAC}, consists of two opposing units, each of our units represented by agents trained using the MARL algorithm. Conversely, enemy units utilize a built-in heuristic algorithm. Our primary objective is to eliminate all enemy units within the specified time frame, so the win rate has become an important evaluation indicator for this environment. In contrast to SMAC, SMACv2 predominantly incorporates random team composition, random initial positions, and aligns the attack range of units with the visible range, thereby enhancing the overall randomness of the environment. In this paper, the version we are utilizing is SC2.4.10. GTDE uses GAT to aggregate intra-group information in this scenario. 

\subsubsection{Battle and Gather}
Battle is a large-scale mixed cooperative competitive scenario where two armies face off against each other in a grid world. Each army consists of 64 agents, and each agent can either move or attack each turn. The goal for both sides is to eliminate the opposing army as quickly as possible. In Gather, agents earn rewards by consuming food. Each piece of food must be broken down by five "attacks" before it can be absorbed. The amount of food on the map is limited, and once a shortage occurs, agents may begin attacking each other in an effort to monopolize the remaining resources. Agents can kill each other with a single attack, and there are a total of 495 agents on the map. In the two scenarios, the observations of each agent are composed of their own features and a circle centered around themselves. GTDE uses matrix multiplication to aggregate intra-group information in this scenario. For Battle and Gather scenarios, we use zero padding observations of dead agents. 

\subsection{Hyperparameter}
GTDE focuses on the actor-critic algorithm, so we select MAAC/MAPPO and IAC/IPPO as representatives of CTDE and DTDE respectively, as our baselines. Compare GTDE based on PPO with MAPPO and IPPO, and compare GTDE based on AC with IAC and MAAC. IAC/IPPO trains the critic using the individual observations of the agent, while MAAC/MAPPO trains a centralized critic using the joint observations of all agents. MAPPO and IPPO are used to evaluate SMACv2, while MAAC and IAC are used to evaluate Battle and Gather. For a fair comparison, all basic hyperparameters are set to be consistent. For SMACv2, Battle, and Gather, we trained 10M, 2000, and 1200 environment steps, respectively. For each scenario, we use 5 different random seeds for all algorithms. Additional detailed hyperparameters are addressed in Appendix \ref{ap:hy}.

\subsection{Performance}
\begin{figure*}[t]
    \centering
    \includegraphics[width=\textwidth]{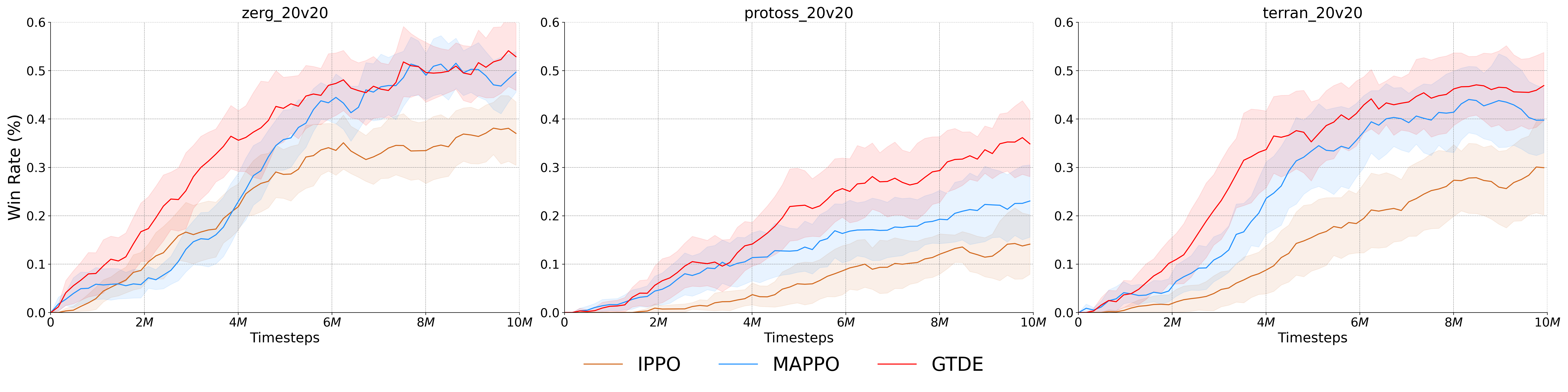}
    \caption{Test curve of average win rate. The shaded area represents the range between the minimum and maximum values across 5 seeds, while the solid line in the center denotes the average value.}
    \label{SMACfig}
\end{figure*} 

\subsubsection{SMACv2}
In the SMACv2 benchmark, since the performances of all algorithms in the 5v5 and 10v10 scenarios are similar, we only report the performance results for the 20v20 scenario. Fig.\ref{SMACfig} shows the test curve of the average win rate. It can be seen that IPPO, which only uses local information, performs the worst in all scenarios. MAPPO and GTDE achieved similar performance in Zerg and Terran scenarios, and better results in the Protoss scenario. This may be due to the lower unit health in Terran and Zerg maps, which leads to shorter episode lengths. In contrast, the presence of shields in the Protoss map can extend the episode duration, giving the agent more time to learn how to effectively utilize local information. As shown in Table \ref{tablewinrate}, we conducted 200 rounds of testing on all models and found that GTDE improved performance by 6.0\% and 17.0\% compared to CTDE and DTDE.

\begin{table}[t]
    \centering
    \begin{tabular}{l|c|c|c|c}
    \toprule
       & zerg & terran & protoss & Battle\\
    \midrule
    GTDE & \textbf{53.4}\scriptsize{(3.2)} & \textbf{48.4}\scriptsize{(2.6)} & \textbf{38.8}\scriptsize{(2.5)} & 52.3\scriptsize{(2.8)} \\
    CTDE & \textbf{50.6}\scriptsize{(1.3)} & \textbf{45.3}\scriptsize{(2.3)} & 26.6\scriptsize{(5.7)} & \textbf{100.0}\scriptsize{(0.0)} \\
    DTDE & 40.8\scriptsize{(1.8)} & 31.3\scriptsize{(2.8)} & 17.6\scriptsize{(5.0)} & \textbf{100.0}\scriptsize{(0.0)} \\
    GTDE-F & \textbf{49.3}\scriptsize{(4.1)} & 42.2\scriptsize{(3.5)} & 32.5\scriptsize{(3.7)} & \textbf{97.2}\scriptsize{(2.6)} \\
    GTDE-U & 46.4\scriptsize{(3.5)} & 43.5\scriptsize{(2.1)} & 32.8\scriptsize{(2.5)} & \textbf{91.8}\scriptsize{(9.9)} \\
    GTDE-A & \textbf{50.3}\scriptsize{(9.6)} & \textbf{46.5}\scriptsize{(0.9)} & \textbf{39.1}\scriptsize{(4.0)} & \textbf{94.5}\scriptsize{(8.3)} \\
    \bottomrule
    \end{tabular}
    \caption{The average win rate and standard deviation were calculated for SMACv2 maps across various models. For the Battle scenario, the win rate represents the outcome of GTDE competing against other algorithms. Values that fall within one standard deviation of the maximum average winning rate are highlighted in bold.}
    \label{tablewinrate}
\end{table}

\subsubsection{Battle}
\begin{figure}[t]
    \centering
    \includegraphics[width=0.9\columnwidth]{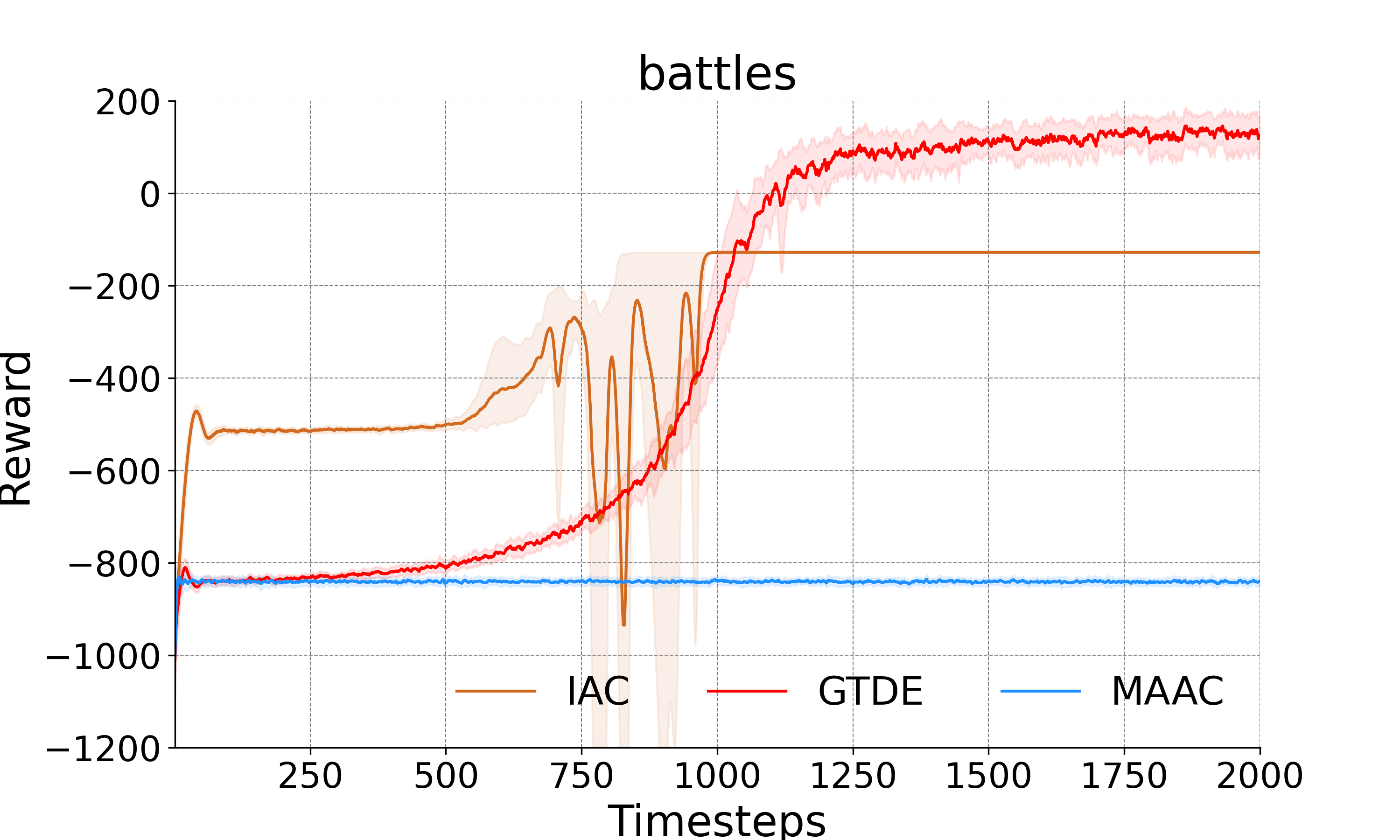}
    \caption{The total reward curve of the Battle scenario(64 v.s. 64).}
    \label{Battlefig}
\end{figure} 

\begin{figure}[t]
    \centering
    \includegraphics[width=0.9\columnwidth]{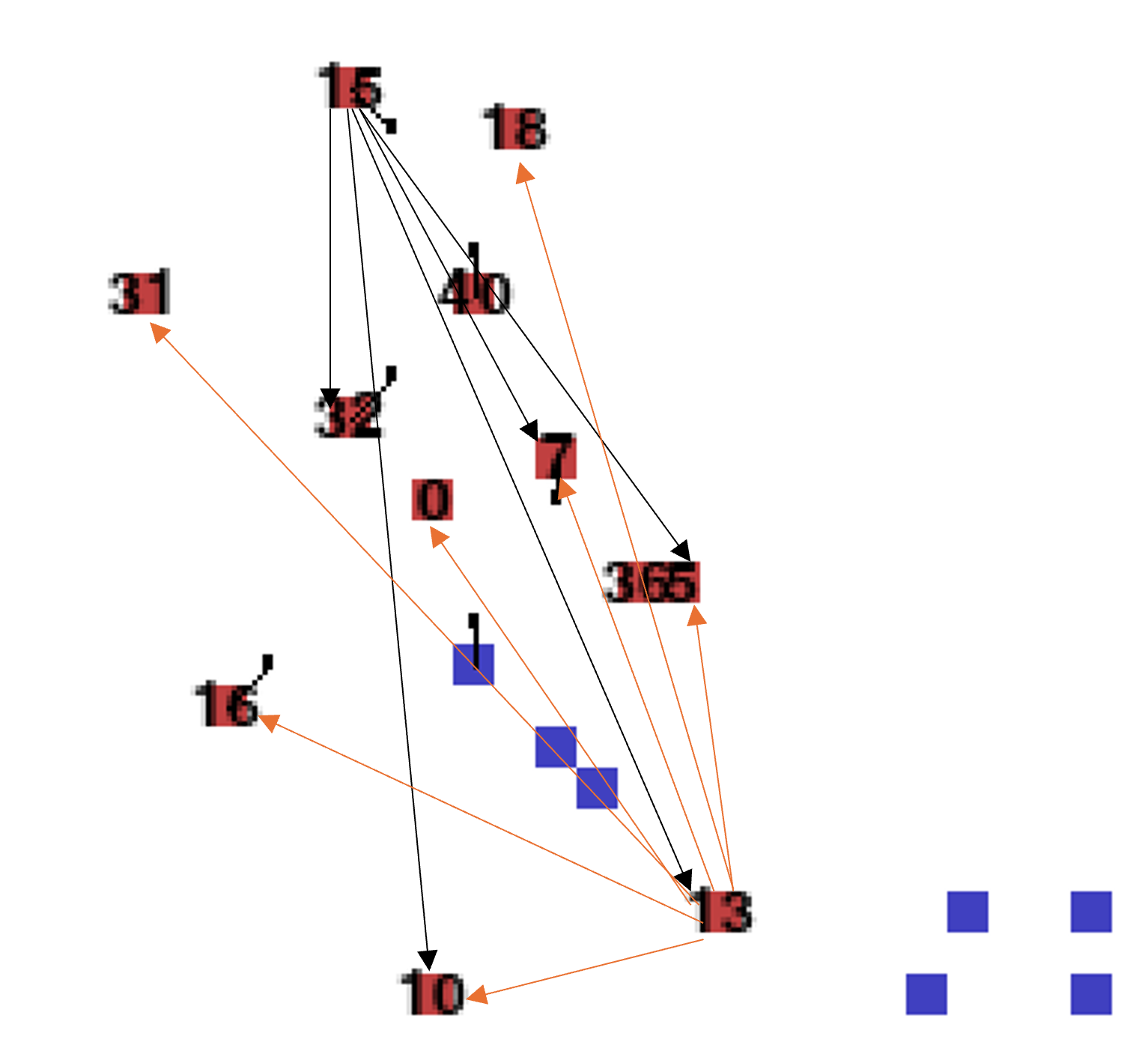}
    \caption{Partial links in the Battle scenario.}
    \label{edge}
\end{figure}

Since there is no built-in heuristic algorithm for the Battle scenario, we employ self-play for training. Fig.\ref{Battlefig} shows the total reward curves of GTDE, MAAC, and IAC. It can be seen that IAC experiences significant fluctuations in learning due to relying solely on its individual observation, ultimately converging to a local optimal solution where neither side attacks. In contrast, MAAC, which requires joint observations of all agents, has a much larger input dimension of $(13\times13\times5+32)\times64$, leading to even lower performance compared to IAC when using the same network size. GTDE, by utilizing local observations instead of individual or joint ones, achieves a more stable improvement in learning performance. Table \ref{tablewinrate} shows the win rate of GTDE against IAC and MAAC algorithms, indicating that GTDE defeated them with a 100\% win rate. Fig.\ref{edge} illustrates the links of partial agents in the Battle scenario. It is evident that Agent 15 is positioned far from the other agents, requiring information to encircle the enemy. Similarly, Agent 13, as part of the besieging group, also depends on information from the other besiegers. 

\subsubsection{Gather}
\begin{figure}[t]
    \centering
    \includegraphics[width=0.9\columnwidth]{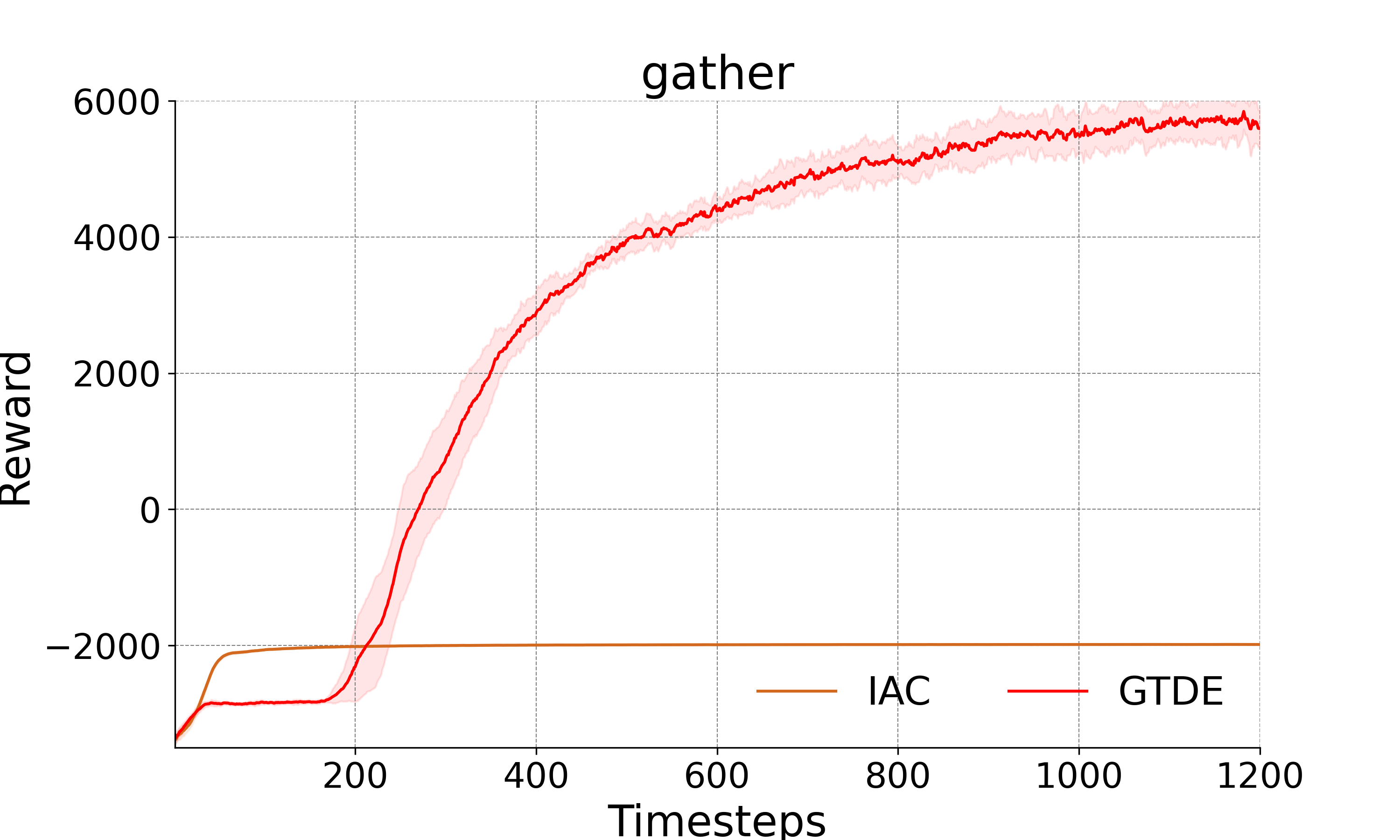}
    \caption{The total reward curve of the Gather scenario(495 agents).}
    \label{Gatherfig}
\end{figure} 

In the gather scenario, the large dimensionality of joint observations demands excessive GPU memory, making it impossible to train MAAC in this scenario. Therefore, Fig.\ref{Gatherfig} only shows the total reward curves of GTDE and IAC. It can be seen that the performance gap between GTDE and IAC has further widened. As the scale of agents increases significantly, centralized training methods become impractical, leaving decentralized training as the only option. However, decentralized methods tend to be less effective. GTDE, which leverages local information, offers a better solution to this issue. 

Furthermore, Table \ref{Information} shows the average agent information used by the three training paradigms in each scenario. Since agents might select agents that have already been eliminated, and these agents do not carry any information(their observation values are all zeros), we exclude dead agents from the table. It can be seen that in three scenarios, GTDE, which uses local information, can achieve similar or better performance compared to models using individual or global information. The average information required for GTDE during training has been reduced by five times, making its training cost in real-world applications much lower than that of CTDE. 

\begin{table}[t]
    \centering
    \begin{tabular}{ccccc}
    \toprule
    Metric & Scenario & GTDE & CTDE & DTDE  \\
    \midrule
    Average node & 20v20 & 4.9\scriptsize{(3.0)} & 20 & 1 \\
    information & Battle & 12.0\scriptsize{(2.0)} & 64 & 1 \\
    ($|g(v_i)|$) & Gather & 103.9\scriptsize{(2.8)} & 495 & 1 \\
    \bottomrule
    \end{tabular} 
    \caption{Agent average information required.}
    \label{Information}
\end{table}

\subsection{Ablations Study}
\begin{figure}[t]
    \centering
    \includegraphics[width=0.9\columnwidth]{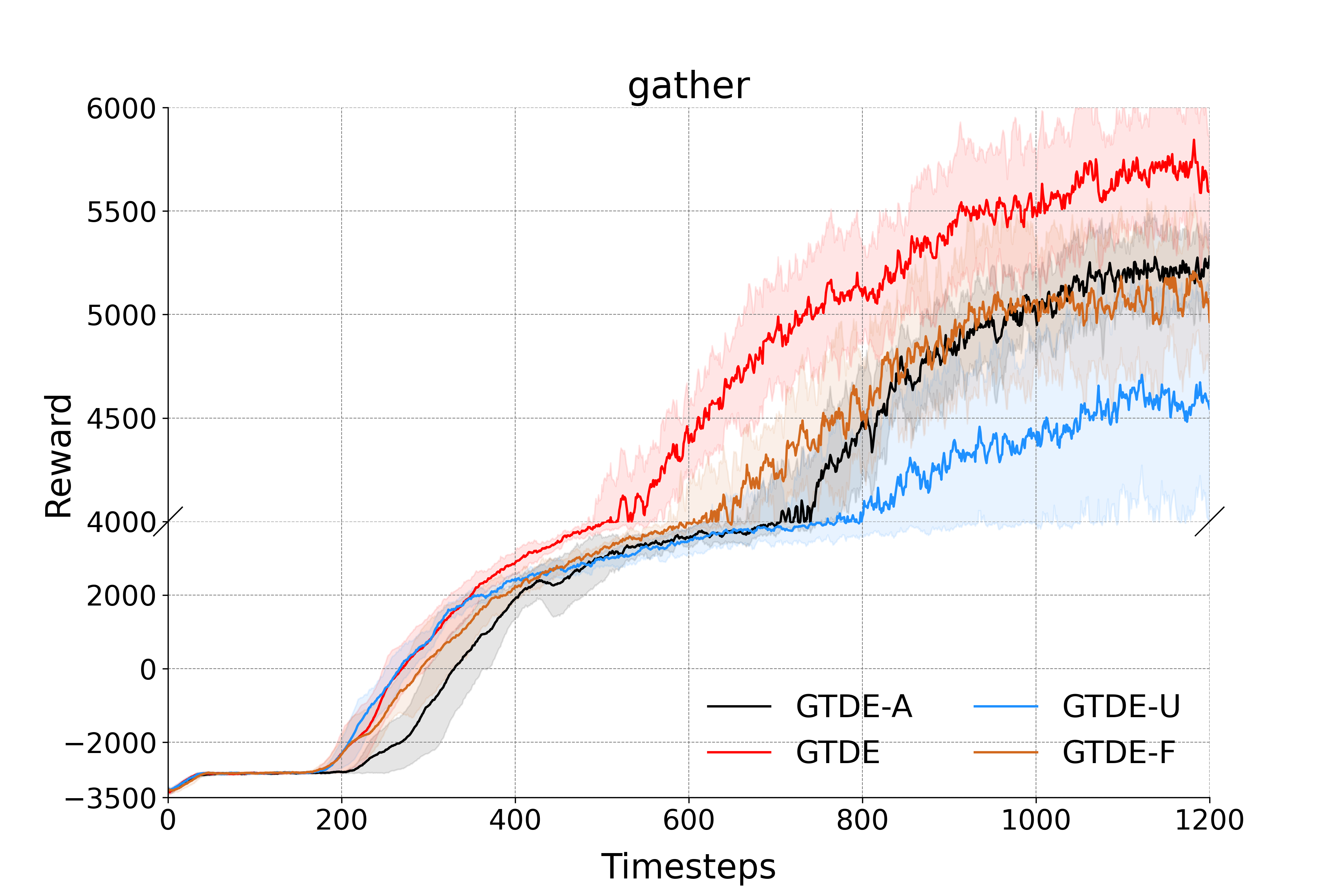}
    \caption{Ablations Study of the Gather scenario.}
    \label{Gatherablfig}
\end{figure} 

We evaluated variants of GTDE to show the effect of the adaptive grouping module. Table \ref{tablewinrate} shows that GTDE outperforms both GTDE-U with random grouping and GTDE-F with fixed grouping, indicating that GTDE learns the correct grouping distribution. In the SMACv2 benchmark, GTDE using local information can achieve performance similar to GTDE-A with pairwise links. In the Battle scenario, all models of GTDE and GTDE-U/A/F underwent 200 rounds of validation and achieved a win rate of over 90\%. Fig.\ref{Gatherablfig} shows the broken axis plot of the total reward in the Gather scenario. We can observe that GTDE received 18.2\%, 11.6\%, and 10.1\% more total rewards than GTDE-U/A/F. Notably, the total rewards of GTDE-F and GTDEA are similar, indicating that using fixed grouping agents can achieve performance comparable to joint observations, which implies that there is information redundancy in joint observations. 

\section{Conclusion and Future Directions}
This paper proposes a GTDE framework to address the limitations of CTDE and DTDE. As the number of agents increases, the performance of DTDE and CTDE significantly deteriorates, with CTDE becoming untrainable due to the high input dimensionality. GTDE is proposed as a solution to this problem. GTDE leverages only the local observations to update policy parameters, making it more suitable for multi-agent systems with a larger number of agents. However, it also faces a challenge in terms of scalability. Unlike CTDE, which uses joint observations and encounters scalability issues due to increasing input dimensions with the number of agents, GTDE avoids this problem but faces a challenge related to the non-scalability of adjacency matrices. It means that new agents can only be linked to existing agents during training and cannot be linked to each other. In the future, we will address or alleviate scalability and train using real-world applications instead of simulators.

\nocite{ddcrp, CoS}
\bibliography{aaai25}

\appendix
\section{Gumbel-Sigmoid} \label{ap:gs}
For the grouping distribution, its adjacency matrix form is as follows:
\begin{equation*}
   \begin{matrix}
      \\
      \text{agent i} \\ 
      \text{agent j} \\
      \text{agent k}
   \end{matrix}
   \underset{(1)}
   {\begin{matrix}
      \begin{matrix}
         i && j && k
      \end{matrix}      \\
      \begin{bmatrix}
         \textbf{0.5} & 0.2 & 0.3 \\ 
         0.1 & \textbf{0.6} & 0.3 \\ 
         \textbf{0.9} & 0.05 & 0.05 
      \end{bmatrix}
   \end{matrix}}
   \ \ \ \ \ \ \ \ \ \ \ \ \
   \underset{(2)}
   {\begin{matrix}
      \begin{matrix}
         i && j && k
      \end{matrix}      \\
      \begin{bmatrix}
         0.3 & \textbf{0.7} & \textbf{0.8} \\ 
         \textbf{0.8} & \textbf{0.6} & \textbf{0.77} \\ 
         0.1 & \textbf{0.6} & 0.45 
      \end{bmatrix}
   \end{matrix}}
\end{equation*}
Each row of the matrix represents a probability distribution derived from the individual observation histories of an agent, where each number in the row denotes the probability($p_{ij}$) of the agent in the $i$-th row links to the agent in the $j$-th column. We need to sample each row of the matrix, where $(1)$ is the adjacency matrix sampled using Gumbel-Softmax. Clearly, it can only sample one. Similarly, Gumbel Top-K can be used to sample row distributions $k$ times. However, the grouping distribution in this paper specifies the number of agents in each group, which is a random variable. Obviously, neither Gumbel-Softmax nor Gumbel Top-K can meet this requirement. Since each element in the matrix follows a Bernoulli distribution, we found that 
\begin{equation*}
   \begin{split}
      Softmax(logp+\epsilon_1)&=\frac{e^{logp+\epsilon_1}}{e^{logp+\epsilon_1}+e^{log(1-p)+\epsilon_2}}\\
                              &=\frac{1}{1+e^{-(log\frac{p}{1-p}+\epsilon_1-\epsilon_2)}}\\
                              &=Sigmoid(log\frac{p}{1-p}+\epsilon_1-\epsilon_2)
   \end{split}
\end{equation*}
where $\epsilon\sim Gumbel(0,1)$. This indicates that using the Bernoulli distribution can represent Gumbel-Softmax via Gumbel-Sigmoid, and each element in the matrix can be efficiently sampled point-to-point by applying the above equation. $(2)$ is the result of applying Gumbel-Sigmoid. Furthermore, we can add masks to ensure that each element in the row is either $0$ or $1$. 

\section{Hyperparameter} \label{ap:hy}
\begin{table}[t]
   \centering
   \begin{tabular}{lc}
   \toprule
   Hyperparameter                      & value           \\
   \midrule
   Num Env Steps                       & 10000000        \\
   Hidden Size                         & 64              \\
   Num Mini Batch                      & 1               \\
   Episode Length                      & 400             \\
   PPO Epoch                           & 5               \\
   Value Loss Coef                     & 1               \\
   Num Rollout Threads                 & 16              \\
   Num training Threads                & 16              \\
   Advantage Calc Method               & GAE             \\
   Use Centralized Value               & False           \\
   Use Value Active Masks              & True            \\
   \hline            
   Multi Head                          & 4               \\
   FC Size                             & 20              \\
   GAT Size                            & 64              \\
   \bottomrule
   \end{tabular}
   \caption{Hyperparameter setting for SMACv2.}
   \label{smacv2table}
\end{table}
For the SMACV2 benchmark, the upper part of Table \ref{smacv2table} displays the basic hyperparameter settings for MAPPO, IPPO, and GTDE, while the lower part presents the additional settings specifically for GTDE.

\begin{table}[!t]
   \centering
   \begin{tabular}{lc}
   \toprule
   Hyperparameter                      & value           \\
   \midrule
   Num Env Steps(Battles)              & 2000            \\
   Num Env Steps(Gather)               & 1200            \\
   Hidden Size                         & 512             \\
   Episode Length                      & 400             \\
   Value Loss Coef(Battles)            & 0.1             \\
   Value Loss Coef(Gather)             & 1               \\
   Entropy Loss Coef(Battles)          & 0.08            \\
   Entropy Loss Coef(Gather)           & 0.01            \\
   $\gamma$(Battles)                   & 0.95            \\
   $\gamma$(Gather)                    & 0.99            \\
   Learning Rate                       & 0.0001          \\
   MiniMap Mode                        & False           \\
   Extra Features                      & False           \\
   Step Reward(Battles)                & -0.005          \\
   Step Reward(Gather)                 & -0.01           \\
   Attack Penalty                      & -0.1            \\
   Dead penalty(Battles)               & -0.1            \\
   Dead penalty(Gather)                & -1              \\
   Attack Opponent Reward              & 0.2             \\
   Attack Food Reward(Gather)          & 0.5             \\
   \hline            
   FC Size(Battles)                    & 64              \\
   FC Size(Gather)                     & 495             \\
   \bottomrule
   \end{tabular}
   \caption{Hyperparameters setting for Battle and Gather.}
   \label{BGtable}
\end{table}
For Battle and Gather scenarios, the upper part of Table \ref{BGtable} displays the basic hyperparameter settings for MAAC, IAC, and GTDE, while the lower part presents the additional settings specifically for GTDE.

\section{Training time} \label{ap:tt}
Regarding training time, the DTDE structure is simple, resulting in an average training time of only 11.8 hours on SMACv2 20v20. The additional overhead in the GTDE framework primarily arises from the extra grouping network and information aggregation module. However, as the grouping and aggregation modules are implemented as small networks, the average training time is 12.8 hours. In contrast, MAPPO has an average training time of 14.6 hours due to its input network size being 20 times larger than that of DTDE and GTDE. 

\end{document}